\documentclass[12pt]{article}
\usepackage{amssymb,amsmath}
\usepackage[noblocks]{authblk}
\usepackage[top=0.75in, bottom=0.75in, left=0.75in, right=0.75in, dvips]{geometry}
\usepackage{caption}
\usepackage{url}
\begin{document}
\textwidth 10.0in
\textheight 9.0in
\topmargin -0.60in
\title{The Canonical Structure of the First Order Einstein-Hilbert Action with a Flat Background}
\author[1,3]{Farrukh Chishtie} \author[1,2]{D.G.C. McKeon}
\affil[1] {Department of Applied Mathematics, The
University of Western Ontario, London, ON N6A 5B7, Canada}
\affil[2] {Department of Mathematics and
Computer Science, Algoma University, Sault St.Marie, ON P6A
2G4, Canada}
\affil[3] {Department of Space Science, Institute of Space Technology, Islamabad 44000, Pakistan}

\maketitle
E-mail: fchishti@uwo.ca, dgmckeo2@uwo.ca

\begin{abstract}
It has been shown that the canonical structure of the first order Einstein-Hilbert (1EH) action involves three generations of constraints and that these can be used to find the generator of a gauge transformation which leaves the action invariant; this transformation is a diffeomorphism with field-dependent gauge function while on shell. In this paper we examine the relationship between the canonical structure of this action and that of the first order spin-2 (1S2) action, which is the weak field limit of the Einstein-Hilbert action. We find that the weak field limit of the Possion Brackets (PB) algebra of first class constraints associated with the 1EH action is not that of the 1S2 action.
\end{abstract}
\noindent
PACS Keywords: Gravity, higher-dimensional, 04.50.-h, Quantum gravity, 04.60.-m, Quantum field theory, 03.70.+k, 11.10.-z, Palatini action, background field, constraint analysis.

\section{Introduction}

An analysis of the 1EH action [1-3] using the Dirac constraint formalism [4,5] reveals the presence of primary, secondary and tertiary constraints. (In an earlier constraint analysis [6,7] of this action, no tertiary constraints arose as equations of motion that are in fact secondary first class constraints were used to eliminate fields from this action.)

The presence of first class tertiary constraints is required to derive a generator of the diffeomorphism transformation. In addition, second class secondary constraints [2] yield unusual ghost contributions to the measure of the path integral used to quantize the 1EH action.

These peculiar features motivate us to examine more clearly the relationship between the 1EH action and the action derived by making a weak field expansion of the metric in the 1EH action about a flat background - the well-known 1S2 action. We find that there is not an obvious connection between the tertiary constraints in these two actions and that the PB algebra of the constraints in the 1EH action does not, in the weak field limit, reduce to the algebra arising in he 1S2 action.

\section{The Constraint Structure}

The 1EH Lagrangian can be written in the form,
\[ \mathcal{L}_{EH} = h^{\mu\nu} \left( G_{\mu\nu ,\lambda}^\lambda + \frac{1}{d-1} G_{\lambda\mu}^\lambda G_{\sigma\nu}^\sigma - G_{\sigma\mu}^\lambda G_{\lambda\nu}^\sigma \right)\eqno(1) \]
in $d$ dimensions, where $h^{\mu \nu}$ and $G^{\lambda}_{\mu \nu}$ are related to the metric $g_{\mu \nu}$ and the affine connection $\Gamma^{\lambda}_{\mu \nu}$ by $h^{\mu \nu}=\sqrt{-g}g^{\mu\nu}$ and
$G^{\lambda}_{\mu \nu}=\Gamma^{\lambda}_{\mu \nu}-\frac{1}{2}\left(\delta^{\lambda}_{\mu}\Gamma^{\sigma}_{\sigma \nu}+\delta^{\lambda}_{\nu}\Gamma^{\sigma}_{\sigma \mu}\right)$.

By making a weak field expansion
\[h^{\mu\nu} (x) = \eta^{\mu \nu} + f^{\mu\nu }(x)\eqno(2)\]

($\eta^{\mu \nu}=diag (- + + + ...)$) and keeping only those terms in $\mathcal{L}_{EH}$ that are bilinear in the fields, one arrives at the first order Lagrangian for a free spin-2 field,
\[ \mathcal{L}_{S2} = f^{\mu\nu} F_{\mu\nu ,\lambda}^\lambda + \eta^{\mu \nu}\left(\frac{1}{d-1} F_{\lambda\mu}^\lambda F_{\sigma\nu}^\sigma - F_{\sigma\mu}^\lambda F_{\lambda\nu}^\sigma \right)\eqno(3) \]

The canonical analysis of both $\mathcal{L}_{EH}$ and $\mathcal{L}_{S2}$ appears in ref. [1] (see also ref. [3] for $\mathcal{L}_{EH}$). This is most easily done for $\mathcal{L}_{EH}$ if one makes the following change of variables

\[ h=h^{00},\quad h^i = h^{0i},\quad  H^{ij} = hh^{ij} - h^i h^j \hspace{1cm}\left(H^{ij}H_{jk}=\delta^i_k\right)\eqno(4-6) \]
\[ G^0_{00} = -\left[\Pi+\frac{\Pi_{ij}}{h}\left(H^{ij}+h^i h^j\right)\right] \eqno(7) \]
\[ G^0_{0i} = -\frac{1}{2}\left[\Pi_i-2\Pi_{ij} h^j\right] \eqno(8) \]
\[ G^0_{ij} = -h\Pi_{ij} \eqno(9) \]
\[ G^i_{jk} = -\xi^i_{jk} \eqno(10) \]
\[ G^i_{0j} = -\frac{1}{2}\left[\overline{\zeta}^i_j-\frac{2}{h}\xi^{i}_{jk}h^k+\overline{t}\delta^i_j\right] (\overline{\zeta}^i_i=0) \eqno(11) \]
\[ G^i_{00} = -\left[\overline{\xi}^i+\frac{1}{h}\xi^{i}_{jk}h^{jk}\right] \eqno(12) \]

One then finds primary constraints that satisfy $\Pi$, $\Pi_i$ and $\Pi_{ij}$ as the canonical momenta conjugate to $h$, $h^{i}$ and $H^{ij}$ respectively. There are subsequently the secondary first class constraints
\[ \chi_i = h_{,i} - h\Pi_{i} \eqno(13) \]
and
\[ \chi = h_{,i}^i + h\Pi  \eqno(14) \]
as well as the secondary second class constraints
\[ \overline{\zeta}_j^i = \frac{2}{h}\left( \lambda_j^i - \frac{1}{d-1} \delta_j^i \lambda_k^k \right)\eqno(15) \]
\[ \xi_{jk}^i = -\frac{1}{2}(M^{-1})_{jk\;\;mn}^{\,i\;\;\;\;\;\;\ell} \sigma_\ell^{mn}\eqno(16)\]
where
\[ \lambda_i^j = h_{,i}^j - \frac{1}{2} h^j \Pi_i - H^{jk}\Pi_{ik}\eqno(17)\]
\[ \sigma_i^{jk} = \frac{1}{h} H_{,i}^{jk} - \frac{1}{h} H^{jk} \Pi_i + \frac{1}{2(d-1)h} \left(\delta_i^j H^{k\ell} + \delta_i^k H^{j\ell}\right) (\Pi_\ell - 2h^m\Pi_{\ell m})\nonumber \]
\[ + \frac{1}{h} \left( h^j H^{kp} + h^k H^{jp}\right)\Pi_{ip} \eqno(18) \]
and
\[(M^{-1})_{yz\;\;\ell m}^{\,x\;\;\;\;\;\;k} = - \frac{h}{2}\bigg[ \left( H_{\ell y} \delta_z^k \delta_m^x + H_{my}\delta_z^k\delta_\ell^x + H_{\ell z} \delta_y^k \delta_m^x + H_{mz} \delta_y^k \delta_\ell^x\right) \eqno(19) \]
\[+ \frac{2}{d-2} \left( H^{k x} H_{\ell m} H_{yz}\right) - H^{kx}
 \left( H_{\ell z} H_{my} +  H_{mz} H_{\ell y}\right)\bigg] .\nonumber \]

Once the second class constraints of eqs. (15,16) have been eliminated through introduction of the Dirac Bracket (DB) the canonical Hamiltonian takes the form
\[\mathcal{H}_c = \frac{1}{h}\left(\tau+h^i\tau_i \right) + F(\chi, \chi_i) \eqno(20)\]

where $F$ is a function, all of whose terms are at least linear in $\chi$ or $\chi_i$, and

\[\hspace{-4cm}\tau = H^{ij}_{,ij} - \frac{1}{2} H_{,n}^{mi}H_{ij} H_{,m}^{nj} - \frac{1}{4} H^{ij} H_{mn,i} H_{,j}^{mn} \eqno(21) \]
\[\hspace{1.4cm} - \frac{1}{4(d-1)} H^{ij} H_{k\ell} H_{,i}^{k\ell} H_{mn} H_{,j}^{mn} + H^{ij} H^{k\ell} \left( \Pi_{ij} \Pi_{k\ell} - \Pi_{ik}\Pi_{j\ell}\right)\nonumber \]
and
\[ \tau_i = -2(H^{mn} \Pi_{mi})_{,n} + H^{mn} \Pi_{mn,i} + (H^{mn} \Pi_{mn})_{,i}.\eqno(22) \]
Since we have the DB algebra
\[ \left\lbrace \chi_i, \chi \right\rbrace = \chi_i \eqno(23)\]
\[ \left\lbrace \tau_i(\vec{x}), \tau_j(\vec{y}) \right\rbrace = \left( -\tau_i (\vec{y}) \partial_j^x + \tau_j(\vec{x})\partial_i^y\right) \delta^{d-1}(\vec{x} - \vec{y})\eqno(24a) \]
\[ \left\lbrace \tau(\vec{x}), \tau(\vec{y}) \right\rbrace = \left( \partial_i^x H^{ij}(\vec{y})\tau_j (\vec{y}) -  \partial_j^y  H^{ij}(\vec{x})\tau_i(\vec{x})\right)\delta^{d-1}(\vec{x}- \vec{y})\eqno(24b) \]
and
\[ \left\lbrace \tau_i(\vec{x}), \tau(\vec{y}) \right\rbrace = \left( - \partial_i^x \tau(\vec{y}) +\partial_i^y \tau(\vec{x})\right) \delta^{d-1}(\vec{x} - \vec{y}).\eqno(24c) \]
we see that there are no further constraints and that ($\chi$, $\chi_i$) are secondary first class constraints and ($\tau$, $\tau_i$) are tertiary first class constraints.

We will now contrast this constraint structure with that which follows from $\mathcal{L}_{S2}$ in eq. (3). In ref. [1] the canonical structure of this 1S2 action was performed, however the variables used there are distinct from those used to analyze $\mathcal{L}_{EH}$ (summarized above). In order to effect a comparison between the canonical structures of $\mathcal{L}_{EH}$ and $\mathcal{L}_{S2}$, we make use of the variables,

\[ F^0_{i j}=\Pi_{ij} \quad F^0_{0 i} = -\frac{1}{2}\Pi_i \quad  F^{0}_{00} = -(\Pi+\Pi_{jj}) \eqno(25a-c) \]
\[ f=1+h \quad f^i = h^{i} \quad  f^{ij} = h\delta^{ij} - H^{ij} \eqno(26a-c) \]
\[ F^{i}_{jk}=-\xi^{i}_{jk} \quad F^i_{00} = -\overline{\xi^i}+\xi^i_{jj} \quad  F_{0j}^i = \frac{1}{2}\left[-\overline{\zeta^i}_j+\overline{t}\delta^i_j \right] (where \overline{\zeta}^i_i=0) \eqno(27a-c) \]

The variables appearing on the right side of eqs. (25-27) are the weak field limit of those in eqs. (4-12) when using the weak limit of eq. (2).

With the change of variable of eqs. (25-27) we find that $\mathcal{L}_{S2}$ becomes
\[ \mathcal{L}_{S2} = \Pi h_{,0} + \Pi_i h_{,0}^i + \Pi_{ij} H_{,0}^{ij} + \frac{d-2}{d-1}\left[(\Pi+\Pi_{ii})^2 - \frac{1}{4}\Pi_i\Pi_i\right] \eqno(28) \]
\[\hspace{1cm} + t\left(h_{,i}^{i}-\Pi \right)+\overline{\xi}^i\left(h_{,i}+\Pi_i\right)+\overline{\zeta}^i_j\left(h^j_{,i}+\Pi_{ij}\right)
+\frac{1}{4}\overline{\zeta}^i_j\overline{\zeta}^j_i \nonumber \]
\[\hspace{1cm} + \xi^i_{jk}\left[-\delta^{jk}\Pi_i - H^{jk}_{,i}+\frac{1}{2(d-1)}\left(\delta^j_i\Pi_k+\delta^k_i\Pi_j\right)\right]+\frac{1}{d-1}\xi^i_{ik}\xi^j_{jk}
-\xi^i_{jk}\xi^j_{ik} \nonumber \]

The equations of motion for $\overline{\zeta}^i_j$ and $\xi^i_{jk}$ clearly constitute a set of secondary second class constraints. If these equations are used to eliminate $\overline{\zeta}^i_j$ and $\xi^i_{jk}$ from $\mathcal{L}_{S2}$, then one can immediately see that the canonical Hamiltonian is given by

\[ \mathcal{H}_{S2} =-\frac{d-2}{d-1}(\Pi+\Pi_{ii})^2 -t\underline{\chi}-\overline{\zeta}^i\underline{\chi}_i+\frac{d-2}{d-1}h^i_{,i}h^j_{,j}+2\Pi_{ij}h^i_{,j} \eqno(29) \]
\[\hspace{1cm}-\frac{2}{d-1}h^i_{,i}\Pi_{jj}+\Pi_{ij}\Pi_{ij} -\frac{1}{d-1}\Pi_{ii}\Pi_{jj}+\frac{1}{4}H^{ij}_{,k}H^{ij}_{,k}\nonumber \]
\[\hspace{1cm}-\frac{1}{4(d-2)}H^{ii}_{,k}H^{jj}_{,k}-\frac{1}{2}H^{ik}_{,j}H^{jk}_{,i}-H^{ij}_{,j}\Pi_i \nonumber \]

where
\[ \underline{\chi}= h_{,i}^i - \Pi,\quad \underline{\chi}_i = h_{,i} + \Pi_{i}. \eqno(30a,b) \]

We see that the momenta conjugate to $t$ and $\overline{\zeta}^i$ vanish; these primary first class constraints lead to the secondary constraints $\underline{\chi}=\underline{\chi_i}=0$. If in eq. (29) we express $\Pi$ and $\Pi_i$ in terms of $\underline{\chi}$ and $\underline{\chi_i}$, then those terms in $\mathcal{H}_{S2}$ that are independent of $\underline{\chi}$ and $\underline{\chi}_i$ are

\[ \mathcal{H}_{S2}^{(0)} =\left(\Pi_{ij}\Pi_{ij}-\Pi_{ii}\Pi_{jj}\right)+2\left(\Pi_{ij}h^i_{,j}-\Pi_{ii}h^j_{,j}\right)+H^{ij}_{,j}h_{,i} \eqno(31) \]
\[\hspace{1cm}-\frac{1}{2}\left(H^{jk}_{,i}H^{ik}_{,j}-\frac{1}{2}H^{ij}_{,k}H^{ij}_{,k}
-\frac{1}{2(d-1)}H^{ii}_{,k}H^{jj}_{,k}\right)\nonumber \]

We find that the secondary constraints $\underline{\chi}$ and $\underline{\chi}_i$ now lead to the tertiary constraints
\[ \left\lbrace \underline{\chi}, \int d^{d-1}\mathcal{H}_{S2}\right\rbrace = -H^{ij}_{,ij} = -\underline{\tau} \eqno(32a)\]
and
\[ \left\lbrace \underline{\chi}_i, \int d^{d-1}\mathcal{H}_{S2} \right\rbrace = 2\left(\frac{d-2}{d-1}\right)\underline{\chi}_{,i} + 2\left(\Pi_{ij,j}-\Pi_{jj,i}\right) \eqno(34b) \]
\[= 2\left(\frac{d-2}{d-1}\right)\underline{\chi}_{,i} + \underline{\tau}_i
 =T_i \nonumber \]

No further constraints arise as

\[ \left\lbrace \underline{\tau}, \int d^{d-1}\mathcal{H}_{S2}\right\rbrace = T_{i,i} \eqno(33a)\]
\[ \left\lbrace \underline{\tau}_i, \int d^{d-1}\mathcal{H}_{S2} \right\rbrace = \underline{\tau}_{,i} + 2\left(\underline{\chi}_{j,i}-\underline{\chi}_{i,j}\right)_{,j} \eqno(33b) \]

Furthermore, ($\underline{\chi}, \underline{\chi}_i, \underline{\tau}, \underline{\tau}_i$) all have vanishing DB amongst themselves. This shows that these constraints are all first class.

Using a technique outlined in ref. [8], (the ``HTZ" approach) it is possible to find the gauge invariances present in an action from the first class constraints that are present. For the $1EH$ action, the first class constraints of eq. (13, 14, 21, 22) have been shown to lead [1, 2] to the diffeomorphism gauge transformation

\[ \delta h^{\mu\nu} = h^{\mu\lambda} \partial_\lambda \theta^\nu + h^{\nu\lambda} \partial_\lambda \theta^\mu -\partial_\lambda (h^{\mu\nu} \theta^\lambda)\eqno(34a) \]

\[ \delta G_{\mu\nu}^\lambda = -\partial_{\mu\nu}^2 \theta^\lambda + \frac{1}{2} \left(\delta_\mu^\lambda \partial_\nu + \delta_\nu^\lambda \partial_\mu\right)\partial \cdot \theta - \theta \cdot \partial G_{\mu\nu}^\lambda  \]
\[ \hspace{1cm}+ G_{\mu\nu}^\rho \partial_\rho \theta^\lambda - \left(G_{\mu\rho}^\lambda \partial_\nu + G_{\nu\rho}^\lambda \partial_\mu\right) \theta^\rho.\eqno(34b)\]

only provided the gauge parameter $\theta^{\mu}$ takes the field dependent form $\theta=-hc$, $\theta^i=c^i-ch^i$ where ($c, c^i$) are arbitrary functions of $x^{\mu}$, and the equation of motion for $H^{ij}$ is satisfied.

For the $1S2$ action, with the first class constraints of eqs. (30, 32), the form of the gauge generator is

\[ G = ap + a_ip_i + b\underline{\chi} +b_i\underline{\chi}_i +c\underline{\tau} + c_i\underline{\tau}_i \eqno(35) \]

Here ($p, p_i$) are the momenta associated with ($t, \overline{\xi}^i$) respectively; these are primary first class constraints. The HTZ formalism shows that in a system with canonical Hamiltonian $H_c$, a set of first class constraints $\phi_{a_{i}}$ (i - generation of the constraint) and total Hamiltonian $H_T=H_c+U_{a_1}\phi_{a_1}$, then the gauge generator $G=\lambda_{a_i} \phi_{a_i}$ satisfies the equation [8]

\[ \frac{D\lambda_{a_i}}{Dt}\phi_{a_i}+{\lbrace G,H_T \rbrace} - \delta U_{a_1}\phi_{a_1} = 0 \eqno(36) \]

where $D/Dt$ denotes the total time derivative exclusive of time dependence through the canonical position and momentum variables. With the generator of eq. (35), eq. (36) leads to

\[ G = \left(-\ddot{c} -\frac{d-3}{d-1}\dot{c}_{i,i}\right)p + \left(\ddot{c}_i-\dot{c}_{,i} + c_{i,jj} - c_{j,ij}\right)p_i + \left(\dot{c}-c_{i,i}\right)\underline{\chi} + \left(-\dot{c}_i+c_{,i}\right)\underline{\chi}_i + c\underline{\tau}+c_i\underline{\tau}_i \eqno(37) \]

From eq. (37), we find that

\[ \delta h=-\dot{c}+c_{i,i},\quad \delta h^i = -\dot{c}_i+c_{,i},\quad  \delta H^{ij} = -c_{i,j}-c_{j,i}+2\delta^{ij}c_{k,k} \eqno(38a-c) \]
Using equation (26), we see that eq. (38) is equivalent to
\[ \delta f^{\mu\nu} = \partial^{\mu}c^{\nu} + \partial^{\nu}c^{\mu} + \eta^{\mu\nu}\partial \cdot c  \eqno(39) \]

where $c^{\mu}=(c, c_i)$. Eq. (39) is the usual spin two gauge transformation; it is the weak field limit of eq. (34a).

We are now in a position to compare and contrast the canonical structure of the $1EH$ and $1S2$ actions. The choice of variables made in eqs. (4-12) has been designed to facilitate this. In particular, if the weak field expansion of eq. (2) is applied in eqs. (4-12) we end up with eqs. (25-27). Furthermore the weak field expansion reduces the secondary first class constraints of the $1EH$ action given by eqs. (13, 14) to the secondary first class constraints of the $1S2$ action given by eqs. (30a, b).

However, beyond this point the canonical analysis of the $1EH$ and $1S2$ actions diverge. We first note that after elimination of the second class constraints, the canonical Hamiltonian for the $1EH$ action is a linear combination of first class constraints (see eq. (20)). This is not the case for the $1S2$ action, as is apparent from eq. (31).

The tertiary constraint $\tau_i$ of eq. (22) in the weak field limit (in which $H^{ij}\approx\delta^{ij}$ reduces to $-2\underline{\tau}_i$ of eq. (32b). However, the weak field limit of $\tau$ given by eq. (21) is not directly related to the constraint $\underline{\tau}$ of eq. (32a). Furthermore, the DB algebra of the first class constraints of the $1EH$ action given by eqs. (23, 24) does not reduce in the weak field limit to the DB algebra of the first class constraints of the $1S2$ action. However, it is surprising that the weak field limit of the gauge transformation associated with the first class constraints following from the $1EH$ action is the gauge transformation for the $1S2$ action following from its first class constraints.

It is apparent that if one were to make the expansion of eq. (2) and substitute it into eq. (1) without dropping terms in the action cubic to the fields, we would have in addition to the $1S2$ action an interaction,

\[ \mathcal{L}_{I} = f^{\mu \nu}\left ({\frac{1}{d-1}F^{\lambda}_{\lambda\nu}}{F^{\sigma}_{\sigma\nu}} - {F^{\lambda}_{\sigma\mu}}{F^{\sigma}_{\lambda\nu}}\right)\eqno(40) \]

Adding this interaction to $\mathcal{L}_{S2}$ clearly does not restore the canonical structure of $\mathcal{L}_{EH}$.

\section{Discussion}

In the preceding section we have demonstrated how expanding the metric about a flat background in the $1EH$ action alters the canonical structure of the theory. If the canonical structure of the 1EH is used in conjunction with the path integral to quantize gravity as in ref. [2], then we are faced with an ambiguous situation when it comes to quantize fluctuations of the gravitational field about a flat background. One could either make use of the expansion of eq. (2) in the path integral of ref. [2], or insert the expansion of eq. (2) into $\mathcal{L}_{EH}$ (eq.(1)) and then derive the path integral that follows from the canonical structure of $\mathcal{L}_{S2}+\mathcal{L}_I$ (eqs. (3, 40)). The two path integrals are not going to be equivalent because they are associated with two distinct canonical structures.

The usual background field method [9, 10] is used in conjunction with the Faddeev-Popov [11, 12] approach to defining the path integral. It has been shown that the use of the background field method is equivalent to what is obtained using canonical quantization when computing radiative corrections in Yang-Mills theory - but not in gravity. However, it has been used to compute loop corrections for the second-order EH (2EH) action [13, 14]. However, from the preceding section it is apparent that using the background field method in conjunction with the path integral as it is derived from the canonical structure of the 1EH action is not equivalent to what is obtained from the 1EH itself. The incompatibility of the path integral derived from the canonical structure of the 1EH action and that which follows from the Faddeev-Popov approach is discussed in ref. [2]; this discrepancy is in part due to the presence of non-trivial ghosts that arise as a result of second class constraints that follow from the 1EH action. Quite likely this inequivalence is also a feature of the path integrals that follow from the 2EH action. (The canonical structure of the 2EH action is discussed in detail in refs. [15-17]). In any case, one should recover the path integral based on the 2EH action from the path integral of the 1EH action by performing the path integral over the affine connection (provided this connection is not coupled to an external source), though the local measure for the path integral could be possibly altered. Problems associated with defining a path integral in systems with non-trivial second class constraints are also discussed in ref [18].

\section*{Acknowledgements}
S. Kuzmin and N. Kiriushcheva had useful comments, as did R.  Macleod.


\begin{thebibliography}{99}
\bibitem{1} D.G.C. McKeon,\textit{Int. J. Mod. Phys.} \textbf{A25}, 3453 (2010).
\bibitem{2} D.G.C. McKeon and F. Chishtie, \textit{Cl. and Quant. Grav.} \textbf{29}  No. 14, 235016 (2012), arxiv 1207.2302.
\bibitem{3} N. Kiriushcheva and S.V. Kuzmin, \textit{Central Eur. J. Phys.} \textbf{9}, 576 (2011).
\bibitem{4} P.A.M. Dirac, ``Lectures on Quantum Mechanics'' (Dover, Minola 2001).
\bibitem{5} M. Henneaux and C. Teitelboim, ``Quantization of Gauge Systems" (Princeton U. Press, Princeton 1992).
\bibitem{6} L.D. Faddeev and V.N. Popov, \textit{Sov. Phys. Usp.} \textbf{16}, 777 (1974).
\bibitem{7} L.D. Faddeev, \textit{Sov. Phys. Usp.} \textbf{25}, 130 (1982).
\bibitem{8} M. Henneaux, C. Teitelboim and J. Zanelli, \textit{Nucl. Phys.} \textbf{B332}, 169 (1990).
\bibitem{9} B.S. DeWitt, \textit{Phys. Rev.} \textbf{162}, 1195 (1967).
\bibitem{10} L. Abbott, \textit{Nucl. Phys.} \textbf{B185}, 189 (1981).
\bibitem{11} L.D. Faddeev and V.N. Popov, \textit{Phys. Lett.} \textbf{B58}, 29 (1967).
\bibitem{12} F. Brandt, J. Frenkel and D.G.C. McKeon, \textit{Phys. Rev.} \textbf{D76}, 105029 (2007).
\bibitem{13} G. 't Hooft and M. Veltman, \textit{Ann. Inst. Henri Poincar$\acute{e}$} \textbf{20}, 69 (1974).
\bibitem{14} M.H. Goroff and A. Sagnotti, \textit{Nucl. Phys.} \textbf{B266}, 709 (1986).
\bibitem{15} N. Kiriushcheva, S. Kuzmin, C. Racknor and S. R. Valluri, \textit{Phys. Lett} \textbf{A372}, 5101 (2008).
\bibitem{16} N. Kiriushcheva and S. Kuzmin, \textit{Cent. Eur. J. Phys.}, 9, 576 (2011).
\bibitem{17} K.R. Green, N. Kiriushcheva and S. Kuzmin, \textit{Cent. Eur. J. Phys.}, \textbf{C71}, 1678 (2011).
\bibitem{18} F. Chishtie and D.G.C. McKeon, \textit{Int. J. Mod. Phys.} \textbf{A27}  No. 14, 1250077 (2012).
\end{thebibliography}
\end{document}